\documentclass[%
 aip,
 amsmath,amssymb,
 reprint,%
]{revtex4-1}

\usepackage[utf8]{inputenc}
\usepackage[T1]{fontenc}
\usepackage{mathptmx}

\usepackage{color}   
\usepackage[normalem]{ulem}   

\usepackage{array}
\usepackage{graphicx}
\graphicspath{{./}{../pdf/}}
\DeclareGraphicsExtensions{.pdf}
\usepackage{dcolumn} 
\usepackage{color}
\usepackage{bm} 

\usepackage{color}

\usepackage{latexsym}

\usepackage[caption=false,font=footnotesize]{subfig}

\begin{document}

\title{Higgs and Goldstone Modes in Crystalline Solids}
\author{Marco Vallone}

\email{e-mail
  \textsf{marco.vallone@polito.it}}

\affiliation{Dipartimento di Elettronica e Telecomunicazioni, Politecnico di Torino, corso Duca degli Abruzzi 24, 10129 Torino, Italy. }

\begin{abstract}
In crystalline solids, the acoustic phonon can be described either as a Goldstone or as a non-Abelian gauge boson. However, the non-Abelianity of the related gauge group apparently makes the acoustic phonon a frequency-gapped mode, in contradiction with the other description. In a different perspective overcoming this contradiction, both acoustic and optical phonon -- the latter never appearing following the other two approaches -- emerge respectively as the gapless Goldstone (phase) and the gapped Higgs (amplitude) fluctuation mode of an order parameter arising from the spontaneous breaking of a global symmetry, without invoking the gauge principle.  In addition, the Higgs mechanism describes all the phonon-phonon interactions, including a possible perturbation of the acoustic phonon's frequency dispersion relation induced by the eventual optical phonon, a peculiar behavior able to produce mini-gaps inside the phonon Brillouin zone.
\end{abstract}

\maketitle


\section{Introduction}
\label{sec:introduction}

The spontaneous breakdown of a continuous symmetry implies the emergence of a massless bosonic particle for each broken generator of the involved symmetry group. This is the Nambu-Goldstone theorem \cite{1960Nambu_PRL,1961Goldstone_JNC} in a nutshell, forming the basis of the Standard Model of the fundamental interactions. In addition, the Higgs mechanism \cite{1964Higgs_PRL}, described as a case of spontaneous symmetry breaking (SSB), plays a prominent role in the Standard Model, providing the mass for the gauge bosons of the electroweak interactions and for the fermions \cite{2010Mandl}. 

These concepts are seemingly far removed from physics of conventional semiconductors and metals commonly employed in electronics and optoelectronics industry. However, the Higgs mechanism, as described by P. W. Higgs in 1964 for particle physics \cite{1964Higgs_PRL}, is the relativistic analog of the plasmon phenomenon described by P. W. Anderson one year before in superconductivity \cite{1963Anderson_PR}, and the existence of Anderson-Higgs modes in condensed matter physics  -- in superconductors, cold-atoms in periodic lattices, in Bose-Einstein condensates, in antiferromagnets, in charge density waves \cite{1981Littlewood_PRL,1982Littlewood_PRB,2011Podolski_PRB} etc.  -- seems pervasive. However, beside plasmons phenomenology \cite{1963Anderson_PR} and related plasmonics \cite{2010Blaber_JPCM,2017Politano_APLMAT}, presently the Higgs mechanism has been acknowledged in semiconductors physics only in somehow exotic materials, like e.g. in topological insulators, Weyl semimetals, cuprates \cite{2010Hasan_RMP,2013Ando_JPSJ,2011Wan_PRB,2011Burkov_PRL,2012Singh_PRB,2015Raines_PRB}, etc., although its role could be much more diffuse even in standard semiconductors. 

On the other hand, the Goldstone theorem is very general and it holds also in the non-relativistic condensed state of matter, where the massless bosons correspond to collective excitations with wavevector $\bm{k}$ and \emph{gapless} frequency dispersion relation $\omega_k$, that is $\omega_{k \rightarrow 0} \rightarrow 0$, where $k = |\bm{k}|$. As a few examples, spin waves in the Heisenberg model are bosons arising when the ground state of the Heisenberg Hamiltonian is magnetically ordered \cite{2005Kendziora_PRL}; collective density excitations in superconductors arise from the spontaneous breaking of the electronic phase rotational $U(1)$ symmetry \cite{2017Pracht,2017Pracht_PRB}. Finally, as a more common example, in crystalline solids, collective excitations associated to lattice vibration modes are the \emph{acoustic} phonons \cite{1976Ashcroft}, and correspond to Goldstone modes emerging from the breaking of a \emph{continuous} spatial symmetry, the translational invariance, broken by the presence of the crystal lattice \cite{1994Leutwyler_PRD,1997Leutwyler_HEPA}. It is said that all these collective excitations originate from one of the so-called \emph{emergence} principles, in this case the Goldstone theorem, since they emerge from the very beginning.

Nevertheless, in condensed matter systems as well as in high energy particle physics, interactions are mediated by \emph{gauge} bosons, appearing when local gauge invariance with respect to a given symmetry group is requested for the system's Lagrangian density \cite{1929Weyl_ZP}. 

In a seemingly similar way, it has been shown \cite{2014Dartora_JPA} that by gauging the spatial translational group $T(3)$ in crystal lattice, three gauge bosons appear to provide the \emph{local} gauge invariance for the Lagrangian under the action of $T(3)$. They can be identified with the three acoustic phonons and, as a major feature of this approach, the elastic properties of solids and the acoustic phonon's dynamical equations can be described in close analogy with General Relativity field equations. In fact, it turns out that the acoustic phonon travels in the crystal acting as a wavelike perturbation of the lattice, similarly to the graviton in vacuum, that travels as a wavelike perturbation of a locally flat differentiable manifold, both obeying very similar field equations \cite{1983Grensing_PRD,1998Gronwald_APPB,2002Low_PRL,2012Watanabe_PRD,2013Blagojevic,2014Brauner_PRD}. Acoustic phonons arise in this case not as Goldstone, but as \emph{gauge} bosons. 

A major concern regards whether the two descriptions could possibly be in contradiction and to what extent. Ref.\,[\onlinecite{2014Dartora_JPA}] describes in detail the \emph{linear limit} of the acoustic phonon's gauge theory, where the two descriptions appear in agreement,  in particular providing in the long wavelength limit ($k \rightarrow 0$) the same gapless  frequency dispersion law $\omega_k = c_s k$, where $c_s$ is the sound velocity in the given medium. However, the cited work does not explore in depth the consequences of the non-Abelianity of the involved gauge group on the ensuing dispersion relation $\omega_k$. 

When more than one ion is present in the lattice elementary cell, another kind of phonon -- the \emph{optical} phonon \cite{1976Ashcroft} -- constitutes a further and independent vibration mode. In polar semiconductors, the electron--longitudinal optical (LO) phonon emission is the dominant intersubband scattering mechanism responsible for electrons thermalization. However, although LO-phonons are of crucial importance in semiconductors transport theory and electron dynamics \cite{1993Sotirelis_PRB,1994Sotirelis_PRB,2014Vallone_NUSOD_my,2015Vallone_PSSB,2017Vallone_JAP}, in the gauge theory of crystal lattice's interactions they are left apart, not arising as gauge bosons.  

In order to clarify these important points, in section\,\ref{sec:phonon_Goldstone} we give a short recap about the emergence of acoustic phonon as Goldstone boson. In section\,\ref{sec:phonon}, after an introduction about  the issues behind gauging a spatial symmetry, we go beyond the linear limit of the theory, finding that the obtained dispersion relation may result gapped (that is, $\omega_{k \rightarrow 0} \neq 0$), in sharp contrast with the gapless dispersion relation characteristic of Goldstone bosons.

In section\,\ref{sec:optical_phonon} we present a different and more general approach, showing that both \emph{acoustic} and \emph{optical} phonons may arise from a SSB which sets out an order parameter $\phi$. It follows that the Higgs mechanism plays an important role: the amplitude fluctuations of $\phi$ are Higgs modes, for which a mass-like term appears in the Lagrangian, that makes the mode gapped, and we identify them with the optical phonons. Conversely, the phase fluctuations of $\phi$ (Goldstone modes) are the acoustic phonons, here revisited in a much more general way that also sheds light on the seeming contradiction arisen when interpreting the acoustic phonon either as Goldstone or as gauge boson. Finally, in section\,\ref{sec:conclusions} main ideas and findings are summarized.

In this work $A^\alpha$ and $B_\beta$ are contravariant and covariant four-vectors, $\partial_\mu$ is the partial derivative $\partial/\partial x^\mu$, where standard Greek indices $\alpha, \beta, \mu,... = 0,... 3$ are indices of space-time coordinates on a four-dimensional differentiable manifold with metric 
$g_{\mu \nu}$ and connection $\Gamma^\alpha_{\mu \nu}$, latin indices ($i, j, k, ... = 1, 2, 3$) mark spatial components, whereas Greek indices with an ``hat'' 
$\hat{\mu}, \hat{\nu},... =0, ...3$ are used for indices of local four-dimensional frames (vierbeins or tetrad indices) on a flat Lorentzian space-time with Minkowski metric 
$\eta_{\hat{\mu} \hat{\nu}} = \mbox{diag}(+1, -1, -1, -1)$. 
The Einstein's summation over repeated index is always understood, and not-italicized ``i'' is the imaginary unit.

\section{Goldstone theorem and acoustic phonons}
\label{sec:phonon_Goldstone}
Apart from superconductivity and exotic materials, things are quite complicated even in the well-known world of solid-state crystals. Let us consider the Lagrangian density
\begin{equation}
\mathcal{L} = \mbox{i} \psi^\dag_{\bm{k}} \partial_t \psi_{\bm{k}} - \frac{1}{2 m^*}  \nabla\psi^\dag_{\bm{k}} \cdot \nabla \psi_{\bm{k}}
\label{eq:Sch_Lagrangian}
\end{equation}
describing the dynamics of a non-relativistic Pauli electron with wavefunction $\psi_{\bm{k}}(t,\bm{r})$ and effective mass $m^*$, free to move in the crystal and obeying the Bloch theorem \cite{1976Ashcroft}. $\mathcal{L}$ is symmetric (i.e. invariant) under global, \emph{continuous} transformations described by the Galilei group $G$ \cite{1994Sternberg,2008Gilmore} (the corresponding relativistic formulation of $\mathcal{L}$ is Lorentz-invariant). If $H$ is the group of time-translations, the generators of the quotient group $G/H$ are the momentum $\bm{p}$ (translations), the angular momentum $\bm{J}$ (rotations), and the boosts $\bm{K}$. Concerning the group of spatial translations $T(3)$ in the ordinary space (one of the subgroups of $G/H$), the elements of the infinitesimal form of $T(3)$ are the operators $U_T = 1 - \mbox{i} \bm{k} \cdot \delta \bm{R}$, where $\delta \bm{R}$ is an infinitesimal displacement of the crystal ions in space around a Bravais lattice \cite{1976Ashcroft} translation vector $\bm{R}$. $U_T$ may be written as $U_T = 1 - \epsilon^j p_j$, where $\epsilon_j$ are real parameters, and the momentum components $p_j = -\mbox{i} \hbar \partial_j$ are the three generators of $T(3)$, whose Lie algebra is described by $[p_{i} , p_{j}] = 0$. Similar considerations could be made for the other two subgroups.

The underlying crystal lattice can be described \cite{1995Peskin} as a potential $-V(\phi)$ to be inserted into $\mathcal{L}$ and eventually depending from several fields $\phi = \{\phi^j\}$. If $\phi_0$ is a the value of $\phi$ that minimizes $V(\phi)$, the system's ground state (the system's true vacuum) is the state for which $\phi = \phi_0$. 
We can expand the potential $V$ around its minimum, obtaining at the second order 
\begin{equation}
V(\phi) = V(\phi_0) + \frac{1}{2}\left(\phi - \phi_0 \right)^k \left(\phi - \phi_0 \right)^j M_{k j} , 
\end{equation}
where 
\begin{equation}
M_{k j} = \left( \frac{\partial^2 V}{\partial \phi^k  \partial \phi^j } \right)_{\phi_0}
\end{equation}
is a symmetric matrix that in the Lagrangian plays the role of a mass term, whose eigenvalues give the eventual masses of the fields $\{\phi^j\}$.

The crystal lattice breaks both translational and rotational invariance making the system's ground state \emph{not symmetric}, and this can happen even if the Lagrangian is symmetric. To be more clear, if we again consider the translations, the fields $\phi^j$ transform under the action of $U_T$ as $\phi^j \rightarrow \phi^j + \alpha(\phi) p^j$, where $\alpha(\phi)$ is an infinitesimal parameter. Nevertheless, the symmetry of the Lagrangian can remain exact, provided 
\begin{equation}
V(\phi^j) = V(\phi^j + \alpha(\phi) p^j) .
\label{eq:Gold0_0}
\end{equation}
Taylor expanding the Eq.\,(\ref{eq:Gold0_0}), the same condition can be written as 
\begin{equation}
\alpha(\phi) \frac{\partial V(\phi)}{\partial \phi^j} = 0 , 
\label{eq:Gold0_1}
\end{equation}
and differentiating the Eq.\,(\ref{eq:Gold0_1}) with respect to $\phi^k$ around the potential minimum $\phi_0$, we obtain
\begin{equation}
\left( \frac{\partial \alpha(\phi)}{\partial \phi^k} \frac{\partial V}{\partial \phi^j}\right)_{\phi_0}  + \alpha(\phi_0) \left( \frac{\partial^2 V}{\partial \phi^k  \partial \phi^j } \right)_{\phi_0} = 0 .
\label{eq:Gold1}
\end{equation}
The first term vanishes, since $\phi_0$ is a minimum of $V$. Regarding the second term, it can be zero if $\alpha(\phi_0) = 0$, but this would be a trivial case: the symmetry would be exact, $V(\phi)$ would be merely an arbitrary constant, it would not be necessary to introduce a field $\phi$, and the crystal lattice simply would not exist. Instead, if $\alpha(\phi_0) \neq 0$, it must be
\begin{equation}
\left( \frac{\partial^2 V}{\partial \phi^k  \partial \phi^j } \right)_{\phi_0} = M_{k j} = 0 ,
\label{eq:Gold2}
\end{equation}
stating that the field is massless.
In summary, the crystal lattice makes a massless field $\phi^j$ to arise, the Goldstone excitation associated to $p^j$. In the end, we expect three Goldstone bosons, one for each broken translations' generators, and the ground state's symmetry is said to be spontaneously broken. It is possible to see that apart from $\bm{p}$, also $\bm{J}$ and $\bm{K}$ are broken generators (i.e. the system's ground state is not symmetric under the corresponding transformations they generate), but it has been shown that they do not give rise to Goldstone bosons \cite{2010Brauner_SYM,2014Brauner_PRD}. Furthermore, it should be remarked that at this level nothing can be said about phonon-phonon interactions.

For uniformity, it is also possible to introduce a pseudo-relativistic notation, describing the acoustic phonons emerged by the Goldstone theorem as sound-like four-momenta $p_\mu$ states in a locally flat manifold $\mathcal{M}$ with metric $g_{\mu \nu}$. The tangent space to any point of $\mathcal{M}$ is a four-dimensional manifold $T_\mathcal{M}$ with the same local \emph{pseudo-Minkowski} metric expression $\eta_{\hat{\mu},\hat{\nu}}$ as in special relativity, and known as \emph{acoustic} or \emph{sound} metric \cite{1981Unruh_PRL,2011Barcelo,2014Dartora_JPA}. Space-time coordinates on $\mathcal{M}$ are expressed as  $x^\mu =  (c_s t, \bm{r})$, and the world ``\emph{pseudo}'' means that the relevant velocity is the sound speed $c_s$  in the considered medium (that of course is not a limit velocity).  

Following this convenient formalism, instead of $T(3)$ we can consider the translations group $T(4)$ with elements
\begin{equation}
U_T = 1 - \epsilon^\mu p_\mu
\end{equation}
on the space-time $\mathcal{M}$, where the parameters $\epsilon^\mu$ are point dependent. The four  $T(4)$ generators are $p_\mu = -\mbox{i}\hbar \partial_\mu$, and the crystal lattice breaks the three spatial components $p_{1...3}$, but not $p_0$. Since $p_\mu p^\mu =0$ (they are sound-like four-momenta in $\mathcal{M}$), a wave-like solution for them yields the \emph{gapless} dispersion relation $\omega_k = c_s k$, valid for small $k$, that makes the Goldstone bosons massless quasi-particles, as required \cite{1988Schrieffer,1999Nagaosa}. The periodic structure of the lattice itself  is known to produce Brillouin-zone folding and the appearance of gaps in the phonon spectrum, i.e., phonon stop bands, for wave vectors satisfying the Bragg condition \cite{1976Ashcroft}. It must be stressed that the present approach does not address this aspect of the problem, since at this level the lattice is treated as a continuum, an approximation valid for wavevectors $k \ll \pi/a$, where $a$ is the lattice constant.

\section{Challenges in gauging spatial symmetry}
\label{sec:phonon}

The existence of Goldstone bosons is closely related to the gauge theories of interactions. The possibility to choose freely a local parameter without changing the physics of a system was declared in 1929 by H. Weyl as a general principle \cite{1929Weyl_ZP}, known as \emph{local gauge invariance}. It turns out that each time a Lagrangian density $\mathcal{L}(\psi_{\bm{k}})$ is requested to be invariant under the action of an element $U_G$ of a Lie group $G$ of \emph{local} transformations $\psi_{\bm{k}} \rightarrow U_G \psi_{\bm{k}}$, one or more compensating fields arise, mediating an interaction (see the description for the classic case of electromagnetism e.g. in Ref.\,[\onlinecite{2010Mandl}], for which the compensating -- or gauge -- field is the vector potential $A_\mu$ that mediates the electromagnetic interaction, and the involved gauge group is the Lie group $U(1)$, whose elements induce phase rotations on $\psi_{\bm{k}}$).

\begin{figure}[!b]
\centerline{\includegraphics[width=0.9\columnwidth]{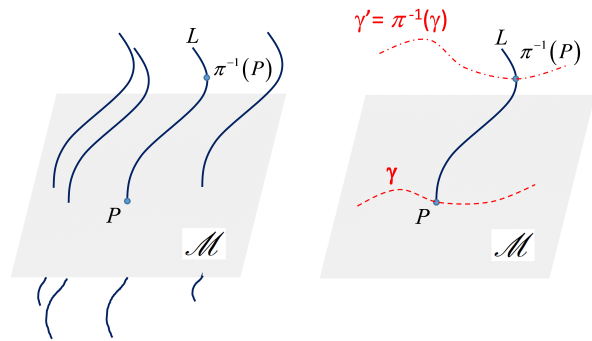}}
\caption{Scheme of a fiber bundle, a map $(\pi :  \mathcal{E} \rightarrow \mathcal{M}, L, G)$, with the base space $\mathcal{M}$ and the fiber $L$, on which the associated group $G$ acts. A trajectory $\gamma$ in $\mathcal{M}$ is mapped to $\gamma\,'$ in the total space $\mathcal{E}$.}
\label{f:1}
\end{figure}

\begin{figure}[!b]
\centerline{\includegraphics[width=0.9\columnwidth]{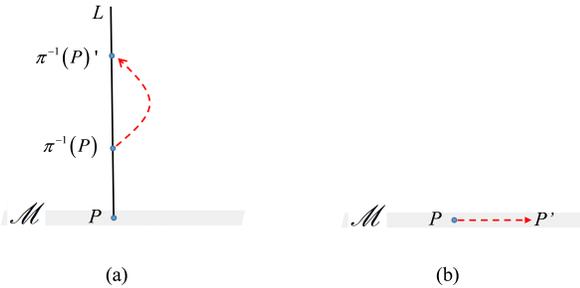}}
\caption{(a) The usual picture of a gauge theory for an internal symmetry: a gauge transformation in $\mathcal{E}$ moves the point $\pi^{-1}(P)$ to $\pi^{-1}(P)'$, preserving the point $P$ on $\mathcal{M}$. (b)  Translations are defined on the base manifold $\mathcal{M}$ itself, where the gauge transformation changes the point $P$ to $P'$.}
\label{f:2}
\end{figure}

In somehow similar way, it is tempting to describe acoustic phonons as gauge fields \cite{2014Dartora_JPA}, like in electromagnetism or, more generally, in Yang-Mills theories \cite{2005tHooft,1980Jackiw_RMP}, considering the \emph{local} (coordinate-dependent) form of $U_T$, that is $U_T = 1 - \epsilon^j(\bm{r}) p_j$. Although $U_T$ is unitary, the Lagrangian $\mathcal{L}$ in the Eq.\,(\ref{eq:Sch_Lagrangian}) is manifestly not invariant under the action of $U_T$ on $\psi_{\bm{k}}$ because of the spatial derivatives in the expression of the generators.
However, the naive pathway to request the Lagrangian $\mathcal{L}$ to be invariant under local spatial translations promoting $T(4)$ to the role of gauge group is not trivial. Considering its importance, before going further it is worth clarifying why gauging a spatial translation is so different from gauging e.g. a wavefunction's phase transformation.
In the gauge theory of electromagnetism, since the involved symmetry transformation is a phase rotation, it is said that an \emph{internal} symmetry is gauged, because the transformation does not involve space-time coordinates. In addition, all this has a simple geometrical meaning: with reference to Figure\,\ref{f:1}, given three differentiable manifolds  $\mathcal{M}$ (the base space), $L(P)$ (the fiber) and $\mathcal{E}$ (the total space), and a Lie group $G$ (the structure group), a fiber bundle is defined as a topological structure $(\pi :  \mathcal{E} \rightarrow \mathcal{M}, L, G)$ where, for each point $P \in \mathcal{M}$ a fiber $L(P)$ is associated to $P$, such that a neighborhood $V_P$ of $P$ is mapped in $\mathcal{E}$ according to $\pi^{-1}(V_P)$, and that $\pi^{-1}(V_P)$ is homeomorphic to the Cartesian product $V_P \times L$. 
The fiber bundle is defined in conjunction with a group $G$ which acts as a transformation group on the fiber, representing the different ways the fiber can be viewed as equivalent. The covariant derivative $D_\mu$ is a \emph{connection} on $\mathcal{E}$ allowing to parallel transport \cite{1980Daniel_RMP,2013Fre} a vector in $\mathcal{E}$. We may also visualize the $n$-dimensional manifold $\mathcal{M}$ as a $n$-surface, with an internal space $L$ associated to each point $P \in \mathcal{M}$ with given topological structure. Specifically, in electromagnetism $\mathcal{M}$ is the ordinary four-dimensional space-time of the special relativity, a phase transformation on $\psi_{\bm{k}}$ is a ``vertical'' automorphism of the bundle and a diffeomorphism of $L$, such that the fiber above each point in $\mathcal{M}$ is unchanged (Figure\,\ref{f:2}(a)). In fact, since only the phase is changing, the point in the total space $\mathcal{E}$ does move, but just ``vertically''  along $L(P)$. This is the gauge freedom carried by  the fiber bundle within its fibers.

By contrast, trying to gauge a space-time transformation group like $T(4)$, it is quite clear that translations in $\mathcal{M}$ induced by $T(4)$ are ``horizontal'' diffeomorphisms of $\mathcal{M}$, since a translation by definition changes the point  $P$ in the base space, involving spatial coordinates. For this reason, an invariance under the action of $T(4)$ is said to be an \emph{external} symmetry. In this context, it is not easy to identify a fiber as an internal space $L$ for which  $\mathcal{E} = \mathcal{M} \times L$ locally holds, preventing a straighforward definition of a fiber bundle $(\pi :  \mathcal{E} \rightarrow \mathcal{M}, L, T(4))$ (Figure\,\ref{f:2}(b)). 
%

\subsection{Gauge theory of spatial translations}
\label{sec:gauging_spatial_translations}

Gauging $T(4)$ in the \emph{pseudo}-Lorentzian, locally flat manifold, equipped with the \emph{acoustic} metric and possibly suitable to describe acoustic phonon dynamics, propagation and interactions, is mathematically not too different from the gauging of $T(4)$ on the standard Lorentzian manifold of the General Relativity \cite{2018Hohmann_PRD}. Nevertheless, the latter revealed a hard goal to obtain \cite{1998Gronwald_APPB,2013Blagojevic}. 
Everything in this field started from the pionieering work by R. Utiyama \cite{1956Utiyama_PR} in 1956, a good starting point, but not entirely convincing, especially because Utiyama did not address the central problem: translations are a diffeomorphism in $\mathcal{M}$, and not in a fiber, as a gauge theory would require, a problem not solved neither by subsequent works by T. Kibble \cite{1961Kibble_JMP} and D. Sciama \cite{1964Sciama_RMP}.
Following the usual Yang-Mills formalism, they defined the covariant derivative as $D_{\hat{\nu}} = \partial_{\hat{\nu}} + W_{\hat{\nu}}$, where $W_{\hat{\nu}}$ is a gauge (compensating) field that, written in terms of its components $R_{\hat{\nu}}^\mu$, reads
\begin{equation}
W_{\hat{\nu}} = R_{\hat{\nu}}^\mu p_\mu .
\label{eq:W_field}
\end{equation}
%
Employing Eq.\,(\ref{eq:W_field}), $D_{\hat{\nu}} $ was written as
\begin{equation}
D_{\hat{\nu}} = h_{\hat{\nu}}^\mu \partial_\mu
\label{eq:covariant_derivative}
\end{equation}
where
\begin{equation}
h_{\hat{\nu}}^\mu = \delta_{\hat{\nu}}^\mu-  \mbox{i} \hbar R_{\hat{\nu}}^\mu .
\label{eq:tetrad}
\end{equation}
The field $h_{\hat{\nu}}^\mu$ defines a set of four orthonormal vectors $h_{\hat{\nu}}^\mu \partial_\mu$, the tetrad or vierbein, provided the covariant index ${\hat{\nu}}$ is a local Lorentz index. 
The correct understanding of $h_{\hat{\nu}}^\mu$ was given by Y. M. Cho \cite{1976Cho_PRD}, who developed a gauge theory of translations with a Yang-Mills--type Lagrangian, where the gauge potentials were correctly interpreted as translational connections (in particular, they are the nontrivial part of the vierbein fields $h_{\hat{\nu}}^\mu$), and not as general coordinate transformations on the base manifold (as e.g. in Utiyama \cite{1956Utiyama_PR}), that would have been not correct. D. and G. Grensing \cite{1983Grensing_PRD} came to similar conclusions, obtaining the gauging of the Poincar{\'e} group in a form that allowed to express General Relativity as a gauge theory of this symmetry group. More recently, the Yang-Mills theory of the \emph{affine} group (the semidirect product
of translations $T(4)$ and general linear transformations $GL(4,R)$) was formulated \cite{2000Tresguerres_PRD,2010Martin_IJGMMP}, where tetrads have been identified with nonlinear translational connections, for which the given $h_{\hat{\nu}}^\mu \partial_\mu$ expression is a simplified yet correct version of the general formulation.  

These concepts were not developed for solid-state physics, but rather to describe gravity, now correctly formulated as a Yang--Mills theory with $T(4)$ as gauge group. Nevertheless, very interestingly the same formulation holds and can be exploited to develop a Yang-Mills theory of elasticity in solid crystals. In fact, Eqs.\, (\ref{eq:W_field}), (\ref{eq:covariant_derivative}) and (\ref{eq:tetrad}) also hold when considering, instead of the standard space-time, the crystal lattice's manifold $\mathcal{M}$ with metric $g_{\mu \nu}$, and with its \emph{pseudo}-Minkowski tangent space $T_\mathcal{M}$ equipped with the sonic metric $\eta_{{\hat{\mu}} {\hat{\nu}}}$. In this case, the field components $R_{\hat{\nu}}^\mu$ 
can be identified with the tensor components of the crystal elasticity field, whereas in theory of gravity they are related to the Ricci's tensor \cite{1972Weinberg}. This allows to formulate the physics of lattice vibrations in crystals (acoustic phonons) in a way that results somehow similar to the physics of gravitational waves in the ordinary space-time, as described in part in Ref.\,[\onlinecite{2014Dartora_JPA}] and developed more in detail here in section\,\ref{sec:coming_to_crystal}.

\subsection{Yang-Mills theory of acoustic phonons in crystal}
\label{sec:coming_to_crystal}

The Lie group $T(4)$ can be promoted to the role of gauge group \cite{2014Dartora_JPA}, defining a gauge-covariant derivative $D_{\hat{\mu}}$ as a connection that makes $\mathcal{L}$ gauge invariant, and consistent with the meaning expressed by Eq.\,(\ref{eq:covariant_derivative}).  
Following the proposed pseudo-relativistic formalism and the standard gauge prescriptions, the infinitesimal displacements of the crystal lattice induce local translations $\psi_{\bm{k}}(x^\mu) \rightarrow \psi'_{\bm{k}}(x^\mu) = U_T(x^\mu) \psi_{\bm{k}}(x^\mu)$ of the electronic wavefunction, where all the ordinary derivatives must be replaced by gauge-covariant derivatives $D_{\hat{\nu}}= h_{\hat{\nu}}^\mu \partial_\mu$. The field $W_{\hat{\nu}}$ must transform in turn according to
\begin{equation}
W_{\hat{\nu}} \rightarrow W'_{\hat{\nu}} = U_T W_{\hat{\nu}} U_T^\dag - \left(\partial_{\hat{\nu}} U_T\right) U_T^\dag ,
\label{eq:field_gauge_law}
\end{equation}
and Eq.\,(\ref{eq:field_gauge_law}) implies 
\begin{equation}
\left(D_{\hat{\nu}}\psi_{\bm{k}}\right)' = U_T \left(D_{\hat{\nu}} \psi_{\bm{k}}\right) ,
\label{eq:field_gauge_law_2}
\end{equation}
that makes the Lagrangian $\mathcal{L}$ gauge invariant under the local action of $T(4)$. 

The translational symmetry manifests itself through the occurrence of conserved currents $ j_\mu$. In the present case, the relevant gauge charges are the three generators $p_k = \int d^3x \,  j_k(x) = \int d^3x \, T_{0 k}(x)$ of the spatial translations, whose corresponding currents are the components of the symmetric energy-momentum tensor $T_{0 k}$. Furthermore, the total energy (Hamiltonian) operator $H = \int d^3x \, c_s^2 \, T_{0 0}(x)$ is written in terms of $p_0$, the unbroken generator of translations in the time coordinate. 

A major difference with respect to electromagnetism is the fact that the elements of $T(4)$ do not commute, even if the group generators on a flat space-time commute, $[p_{\hat{\mu}}, p_{\hat{\nu}}] = 0$. In fact, if $U_T(x)$ and $U_T(y)$ are two elements of $T(4)$, it is 
\begin{equation}
[U_T(x), U_T(y)] = -\hbar^2 \left[\epsilon^\mu(x)  \partial_\mu \epsilon^\nu(y)  - \epsilon^\mu(y) \partial_\mu\epsilon^\nu(x)   \right]  \partial_\nu ,
\label{eq:commuting_rules}
\end{equation}
that in general is nonzero for $x \neq y$, making non-Abelian the local form of the gauge group. The intrinsic difference between the structure of the global and local versions of $T(4)$ can be stated more formally writing the currents algebra
\begin{equation}
[ j_\mu(x), j_\nu(y)] = - \mbox{i} \hbar \left( \partial_\mu j_\nu(y) -  \partial_\nu j_\mu(x) \right) \delta^3\left(x-y\right) ,
\label{eq:commuting_rules2}
\end{equation}
where the commutator in general is nonzero for $x \neq y$. Hence the generators' algebra can be written as \cite{2016Gegenberg_PRD}
\begin{equation}
[p_\mu , p_\nu] = \mbox{i} g f^\gamma _{\mu \nu} p_\gamma 
\label{eq:commuting_rules3}
\end{equation}
where $f^\gamma _{\mu \nu}$ are the structure constants for the currents' Lie algebra and $g$ is the coupling constant of the theory. It is $f^\gamma _{\mu \mu} = 0$, and $f^\gamma _{\mu \nu} = - f^\gamma _{\nu \mu} = 1$ for $\mu \neq \nu$. In a flat space-time $T_\mathcal{M}$ of course it is $[p_{\hat{\mu}} , p_{\hat{\nu}}]  = 0$, but this is not true on $\mathcal{M}$.
This is a central point of the theory, and the algebra expressed by Eq.\,(\ref{eq:commuting_rules3}) makes non-Abelian the local symmetry, and it is responsible for the phonon-phonon interactions. On this basis, it also follows
\begin{equation}
[W_{\hat{\mu}}, W_{\hat{\nu}}] = \mbox{i} g f^\alpha _{\beta \gamma}  R_{\hat{\mu}}^\beta R_{\hat{\nu}}^\gamma p_\alpha
\label{eq:commutator_W}
\end{equation}
Making use of Eq.\,(\ref{eq:covariant_derivative}), the commutator $[D_{\hat{\mu}}, D_{\hat{\nu}}]$ can be written in the alternative forms
\begin{equation}
[D_{\hat{\mu}}, D_{\hat{\nu}}] = G_{{\hat{\mu}} {\hat{\nu}}}^\alpha p_\alpha  = G_{{\hat{\mu}} {\hat{\nu}}}
\label{eq:commutator}
\end{equation}
where the field strength tensor $G_{{\hat{\mu}} {\hat{\nu}}}$ and its components $G_{{\hat{\mu}} {\hat{\nu}}}^\alpha $ are given by
\begin{eqnarray}
G_{{\hat{\mu}} {\hat{\nu}}} =&& \partial_{\hat{\mu}} W_{\hat{\nu}} - \partial_{\hat{\nu}} W_{\hat{\mu}} + [W_{\hat{\mu}}, W_{\hat{\nu}}]  \nonumber \\
G_{{\hat{\mu}} {\hat{\nu}}}^\alpha =&& \partial_{\hat{\mu}} R_{\hat{\nu}}^\alpha - \partial_{\hat{\nu}} R_{\hat{\mu}}^\alpha +\mbox{i} g f^\alpha_{\beta \gamma}  R_{\hat{\mu}}^\beta R_{\hat{\nu}}^\gamma .
\label{eq:field_strength_tensor}
\end{eqnarray}
It is worth noting that Eqs.\,(\ref{eq:field_strength_tensor}) resembles expressions typical of \emph{classical} Yang-Mills theories, although it should not be forgotten that the symmetry group rules an external symmetry, and the gauge field $W_{\hat{\mu}}$ contains spatial partial derivatives, since the infinitesimal group generators act as differential operators. 

The Lagrangian density for the free gauge field is 
\begin{equation}
\mathcal{L}_{R} = \frac{1}{4} G_{{\hat{\mu}}{\hat{\nu}}}^\alpha G^{{\hat{\mu}} {\hat{\nu}}}_\alpha ,
\end{equation}
that explicitly reads
\begin{eqnarray}
\mathcal{L}_{R} =&&\frac{1}{2} \left(\partial_{\hat{\mu}} R_{\hat{\nu}}^\alpha \partial^{\hat{\mu}} R^{\hat{\nu}}_\alpha  -  \partial_{\hat{\mu}} R_{\hat{\nu}}^\alpha \partial^{\hat{\nu}} R^{\hat{\mu}}_\alpha \right)   \nonumber \\
           &&+ \mbox{i} g f_\alpha^{ \gamma \delta}\,\, R^{\hat{\mu}}_\gamma R^{\hat{\nu}}_\delta \, \left(\partial_{\hat{\mu}} R_{\hat{\nu}}^\alpha \right)  \nonumber \\
           &&- \frac{g^2}{4} f^\sigma_{\beta \gamma} f_\sigma^{\delta \epsilon} R_{\hat{\mu}}^\beta R_{\hat{\nu}}^\gamma R^{\hat{\mu}}_\delta R^{\hat{\nu}}_\epsilon
\label{eq:Lagrangian_complete}
\end{eqnarray}
and describes the dynamics of a field with cubic and quartic (self-interacting) terms. Writing the Eulero-Lagrange equation 
\begin{equation}
\partial^{\hat{\mu}}  \frac{\partial\mathcal{L}_R}{\partial\left(\partial^{\hat{\mu}} R^{\hat{\nu}}_\alpha\right)} = \frac{\partial\mathcal{L}_R}{\partial R^{\hat{\nu}}_\alpha}  
\end{equation}
and imposing the Lorenz gauge (that allows for great simplification of all the expressions), the equation of motion for the free field $R_{\hat{\nu}}^\alpha$ results (defining $\lambda = g^2$ and $\Box = \partial_{\hat{\mu}}\partial^{\hat{\mu}}$)
\begin{equation}
\Box  R_{\hat{\nu}}^\alpha + \lambda \,  f^\sigma_{\beta \gamma} f_\sigma^{\delta \alpha} R_{\hat{\mu}}^\beta R_{\hat{\nu}}^\gamma R^{\hat{\mu}}_\delta  = 0 . 
\end{equation}
This equation does not include any mass-term (it would be a term linear in $R_{\hat{\nu}}^\alpha$), but a cubic self-interacting term is present, coming from the quartic term in the Lagrangian. 

All the components are highly coupled and difficult to manage. In order to understand the underlying physics, it is particularly interesting to consider in better detail the case of a simple one-dimensional atomic chain along $x$, assuming -- without loosing generality -- that the perturbation is along the longitudinal direction (longitudinal acoustic phonon), hence $R_0^0 = 0$. From Eq.\,(\ref{eq:commuting_rules3}), that expresses the group commutation rules, the motion equations for the nonzero field components $\hat{R} = R_0^1$ (the longitudinal \emph{acoustic phonon} field) and $\hat{S} = R_1^1$ (the \emph{spatial strain} field) decouple, yielding
\begin{equation}
\left( \partial^2_t  - c_s^2 \partial^2_x \right) \hat{R} + \lambda c_s^2  |\hat{R}|^2 \hat{R} = 0   
\label{eq:acousticphonons_motioneq}
\end{equation}
\begin{equation}
\left( \partial^2_t  - c_s^2 \partial^2_x \right) \hat{S} = 0   . 
\label{eq:strain_motioneq}
\end{equation}
$\hat{S}(x,t)$ describes the fluctuations of the crystal strain field, that in higher dimensional domains are coupled to phonons, which may get scattered by them. Regarding the equation for the acoustic phonon $\hat{R}(x,t)$, the parameter $\lambda$ leads to a nonlinear oscillatory solution for $\hat{R}$ with a dispersion relation given by \cite{2011Frasca_JNMP,2017Frasca_EPJP} 
\begin{equation}
\omega_k^2 = \sqrt{\frac{\lambda c_s^2 }{2}} \rho_R^2 +  c_s^2 k^2
\label{eq:dispersion_acoustic}
\end{equation}
where $\rho_R$ is an integration constant depending on the cell characteristics. Eq.\,(\ref{eq:acousticphonons_motioneq}) can be derived by the Eulero-Lagrange equation from a Lagrangian that is a simplified form of Eq.\,(\ref{eq:Lagrangian_complete}),
\begin{equation}
\mathcal{L}_R = \frac{1}{2} \left(\eta^{\hat{\mu}\hat{\nu}} \partial_{\hat{\mu}}R \, \partial_{\hat{\nu}}R - \frac{1}{2}\lambda R^4 \right) ,
\label{eq:quartic_R}
\end{equation}
the well-known Lagrangian with $\phi^4$-potential, describing a massless self-interacting scalar field \cite{2010Mandl}.  
Before concluding this section, we recap the path followed so far: the local gauge invariance imposed for the electron's Lagrangian in Eq.\,(\ref{eq:Sch_Lagrangian}) makes three gauge fields to arise, identified with acoustic phonon modes, one longitudinal and two transverse. Their Lagrangian is the Eq.\,(\ref{eq:Lagrangian_complete}), for which a simplified, one-dimensional form is the Eq.\,(\ref{eq:quartic_R}), describing phonon-phonon interactions through the nonlinear (quartic) term proportional to $\lambda$.   

It is also worth stressing that Eq.\,(\ref{eq:dispersion_acoustic}) describes a possible frequency gap $\sqrt{\lambda c_s^2/2} \rho_R^2$ at the long-wavelength side of the spectrum. This gap, not described by the classical theory of acoustic phonons, is due to the fields' self-interaction term coming from the nonzero commutators in Eq.\,(\ref{eq:commutator_W}) and Eq.\,(\ref{eq:field_strength_tensor}), in turn originating from the current algebra in Eq.\,(\ref{eq:commuting_rules2}) and Eq.\,(\ref{eq:commuting_rules3}). However, since the coupling constant can be very small, the frequency (energy) gap may be a very tiny mini-gap, probably beyond practical observability limit. Moreover, it is also worth remarking that the present gauge theory is a massless, non-Abelian gauge theory, not a massive one. Apart from issues in its renormalizability, a mass-term in Eq.\,(\ref{eq:quartic_R}) would be a term proportional to $R^2$, not existent in the present formulation, beside the quartic term coming from the nonlinearities. Nevertheless, the final result in the phonon dispersion relation would be similar (a mini-gap in the spectrum), but its very origin would be very different: a massive boson, and not a self-interacting massless boson as in the present case. 

In summary, we may say that in solid crystals a Goldstone mode is a massless excitation, which very existence is due to the broken, continuous translational symmetry caused by the lattice, and its dynamics can be described in the framework of a non-Abelian gauge theory. It is worth considering that Eq.\,(\ref{eq:dispersion_acoustic}) reduces to the well known Debye form\cite{1976Ashcroft} $\omega_k =  c_s |k|$, provided the nonlinear term in the Lagrangian can be neglected. Nevertheless, a relevant and quite unexpected result is that, in its general form, $\omega_k$ results \emph{gapped} (i.e. $\omega_k \neq 0$ in the limit $k \rightarrow 0$) owing to self-interactions, if the latter are relevant, even though the gauge field is a massless boson. 

As a side feature already noticed in Ref.\,[\onlinecite{2014Dartora_JPA}], the commutator $[D_{\hat{\mu}}, D_{\hat{\nu}}]$ satisfies a cyclic identity that can be written as
\begin{equation}
D_\sigma G_{\hat{\mu}\hat{\nu}} + D_{\hat{\nu}}G_{\hat{\sigma} \hat{\mu}} + D_{\hat{\mu}}G_{\hat{\nu}\hat{\sigma}} = 0  , 
\label{eq:Bianchi_G}
\end{equation}
an equation known as the Bianchi identity for the field strenght tensor. The tensor $R_{\hat{\mu}}^\nu$ plays the same role of the Ricci tensor in General Relativity, obeying the same dynamic equations implied by Eq.\,(\ref{eq:Bianchi_G}), and describing the local perturbation of the sonic metric $\eta_{\hat{\mu} \hat{\nu}}$ of the solid crystal caused by the propagation of an acoustic phonon, in strict analogy with the space-time Lorentz metric perturbation induced by a gravitational wave propagating as described in General Relativity. 

\section{Optical phonon: a Higgs mode}
\label{sec:optical_phonon}

\begin{figure}[!t]
\centerline{\includegraphics[width=0.9\columnwidth]{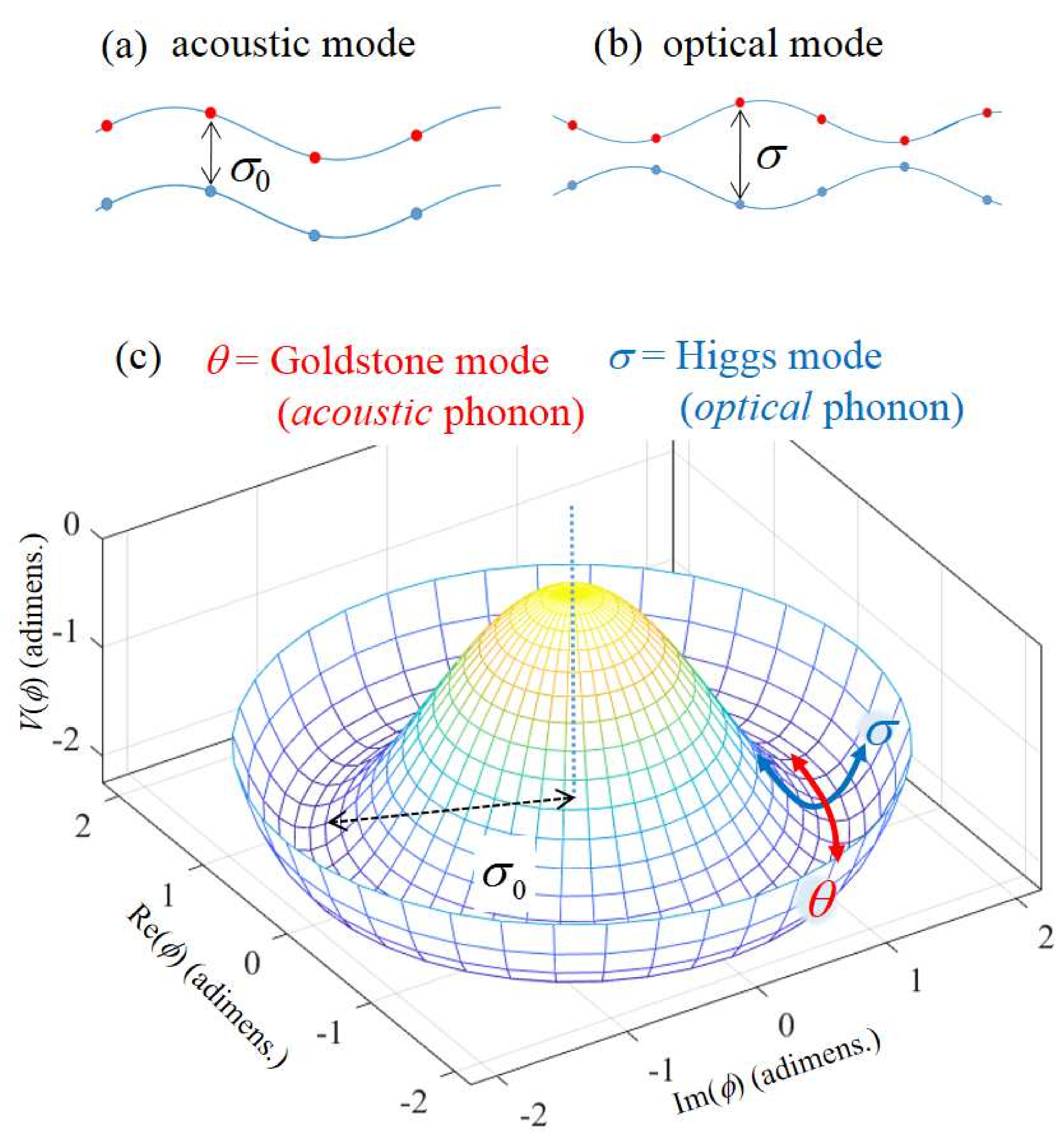}}
\caption{Scheme of a diatomic one-dimensional chain showing the in-phase (a) and out-of-phase (b) oscillating modes. (c) The Mexican-hat potential $V(\phi)$, leading to the excitation of the amplitude (or Higgs) and phase (or Goldstone) modes, $\sigma$ and $\theta$. }
\label{f:3}
\end{figure}

In several binary compounds of interest in electronics, as e.g. GaAs, InP, GaN and many others, the crystal can be considered as composed of two interacting sublattices, one for each atomic species. Textbook toy-models describe the relevant physics considering a one-dimensional Bravais lattice with two ions per primitive cell. When treated as a set of elementary classical oscillators, they show that two distinct oscillation modes exist \cite{1976Ashcroft}: 

$a$) the two atoms in the cell may oscillate at frequency $\omega_k$ around their equilibrium position with the same direction and phase (Figure\,\ref{f:3}(a)), keeping unchanged their relative displacement $\sigma_0$. The perturbation is  the \emph{acoustic} mode, an elastic wave propagating in the lattice at the sound velocity $c_s$, and the corresponding collective ions excitation is the acoustic phonon, whose dispersion relation has an expression that in the long wavelength limit can be approximated as $\omega_k = c_s |k|$, indicating that the acoustic mode can be excited with arbitrarily small energy (there is no gap in $\omega_k$);

$b$) conversely, the two atoms in the cell may oscillate opposite to each other (Figure\,\ref{f:3}(b)), and their relative displacement $\sigma$ oscillates around an equilibrium position $\sigma_0$ along a well-defined direction $\theta_0$ respect to a reference, defining an oscillating cell's dipole moment, if the solid is ionic. This is the well known \emph{optical} mode, and the associated collective excitation is the \emph{optical phonon}, whose role in semiconductors transport theory and electron dynamics is of great importance. It is found that $\omega_{k \rightarrow 0} \rightarrow \omega_0 \neq 0$: in other words, the optical phonon dispersion $\omega_k$ is \emph{gapped}, and the optical mode's excitation costs a fixed nonzero energy $\hbar \omega_0$ also for $k \rightarrow 0$. 

We have found that both acoustic and optical phonons may be described as arising from a spontaneous breaking of a global symmetry, and that they are respectively a Goldstone (phase) and a Higgs (amplitude) mode \cite{1960GellMann_JNC,2011Podolski_PRB,2013Nicolis_PRL} of an order parameter oscillation. Beside its importance as a further sign of a widespread presence of the amplitude mode in condensed-matter systems with broken continuous symmetry, the present approach provides a unified formulation of phonons, where the coupling between the acoustic and the optical phonon modes arises in a natural way.

In order to show this important point, let us define the displacement between the two atoms in the cell as a complex field $\phi = \left( \sigma_0+\sigma \right) \exp\left(\mbox{i} \theta \right)$.  
We can describe the dynamics of the collective oscillation modes by the Lagrangian of the field $\phi$  
\begin{equation}
\mathcal{L}_{\phi} = \frac{1}{2} \left(\partial_{\hat{\mu}}\phi \, \partial^{\hat{\mu}}\phi^* - V(\phi) \right) ,
\label{eq:quartic}
\end{equation}
where $V(\phi)$ is the interacting potential between the two atoms in the cell. 
With the choice $V(\phi) = (1/4) \lambda |\phi|^4$, we would obtain the Eq.\,(\ref{eq:quartic_R}), able to describe acoustic phonons. We could assume for $V(\phi)$ the form 
\begin{equation}
V(\phi) =  \frac{1}{2} \frac{\mu^2 c_s^2}{\hbar^2} |\phi|^2+ \frac{1}{4}\lambda |\phi|^4 ,
\label{eq:V_sigma_model}
\end{equation}
typical of the $\phi^4$ theories \cite{2010Mandl} with a mass term proportional to $\phi^2$ . However, if the two parameters $\mu^2$ (where $\mu$ has the dimension of a mass) and $\lambda$ are both positive, the expectation value of $\phi$ in the physical vacuum state is $<\phi> = 0$ (the minimum of $V(\phi)$). 
In this case the system is in a disordered phase, since there is not a preferred direction for the atoms' displacement oscillation. 
Furthermore, the system's vacuum energy is invariant under a $U(1)$ transformation, since $\mathcal{L}_{\phi}$ is invariant under a \emph{global} $U(1)$ transformation of the field $\phi$. 
Consequenty, this model cannot describe optical phonons, for which $\theta = \theta_0$ results a \emph{preferred} direction for the system. 
In addition, the introduction of the mass term is arbitrary, beside making the theory non-renormalizable. Furthermore, the theory would describe massive acoustic phonons, against all evidences.

Instead, observing that the optical phonon defines an oscillating cell's dipole moment, we may inspire us to ferromagnets, assuming that the global phonon field might have a preferred direction, violating a symmetry of the Lagrangian. 
In this case, the related field theory has a continuous, hidden symmetry that is spontaneously broken by a term in the Lagrangian that cannot be derived, but only inserted in $V(\phi)$ in an \emph{ad hoc} manner, as in the electroweak theory \cite{2010Mandl,1995Peskin}. 
To this end, we proceed in the standard way according to the linear-sigma model \cite{1960GellMann_JNC,2013Nicolis_PRL}. If the crystal formation makes $\mu^2$ negative, the potential $V(\phi)$ assumes the well known ``Mexican-hat''  shape shown in Figure\,\ref{f:3}(c), and all the couples $(\sigma, \theta)$ on the circle with radius $|\phi| = \sqrt{-\mu^2 c_s^2/(\hbar^2 \lambda)} = \sigma_0$ are $V(\phi)$ minima. They identify degenerate vacuum states, all with the same energy, and among them the system chooses the \emph{ordered state} for which $<\phi> = (\sigma_0, \theta_0)$, that becomes the true vacuum state of the system. 
The angle $\theta_0$ identifies the \emph{preferred} direction imposed by the cell structure and, without loosing generality, we can set $\theta_0=0$ as reference $x$-axis. Hence, the Lagrangian in Eq.\, (\ref{eq:quartic}) with the potential in Eq.\,(\ref{eq:V_sigma_model}) and negative $\mu^2$ has a spontaneously broken symmetry. 

The next important step is to write the Lagrangian $\mathcal{L}_{\phi}$ in terms of the fields $\sigma$ and $\theta$, small fluctuations of $\phi=(\sigma_0 + \sigma,\theta)$ around the true vacuum $\phi_0 = (\sigma_0, 0)$:
\begin{equation}
\mathcal{L}_{\phi} = \mathcal{L}_{\theta} + \mathcal{L}_{\sigma} + \mathcal{L}_{\sigma-\theta} ,
\label{eq:quartic_2_0}
\end{equation}
where
\begin{eqnarray}
\mathcal{L}_{\theta} =&&  \frac{1}{2} \partial_{\hat{\mu}}\theta \, \partial^{\hat{\mu}}\theta - \frac{\lambda \theta^4 }{4}  \label{eq:quartic_2_theta} \\
\mathcal{L}_{\sigma} =&&  \frac{1}{2}\left( \partial_{\hat{\mu}}\sigma \, \partial^{\hat{\mu}}\sigma  -  \frac{\omega_0^2}{c_s^2} \sigma^2  \right)  - \sqrt{\frac{\lambda \omega_0^2 }{2 c_s^2}}  \sigma^3  - \frac{\lambda \sigma^4 }{4}  \label{eq:quartic_2_sigma} \\
\mathcal{L}_{\sigma-\theta} =&&  - \sqrt{\frac{\lambda \omega_0^2 }{2 c_s^2}}  \sigma \theta^2 -  \frac{\lambda}{2} \theta^2\sigma^2 \, .
\label{eq:quartic_2_sigma_theta}
\end{eqnarray}
Here, for later convenience, we have introduced the real and positively defined parameter $\omega_0^2 = -2\mu^2 c_s^4/\hbar^2$, a factor playing the dynamical role of a mass-term. The Lagrangians $\mathcal{L}_{\theta}$ and $\mathcal{L}_{\sigma}$ respectively describe the dynamics of fields $\theta$ and $\sigma$. As expected and as it is typical for the Higgs' mechanism, $\theta$ is massless (Eq.\,(\ref{eq:quartic_2_theta}) does not include a term proportional to $\omega_0^2$), whereas in Eq.\,(\ref{eq:quartic_2_sigma}) a mass-term with the correct sign is associated to the field $\sigma$. The Lagrangian $\mathcal{L}_{\sigma-\theta}$ describes the $\sigma-\theta$ coupling terms and will be treated in section\,\ref{sec:phonons-coupling}.

It is remarkable that the vacuum state is no more $U(1)$ invariant (it is non-symmetric in $\theta$), although Eq.\,(\ref{eq:quartic}) and Eq.\,(\ref{eq:quartic_2_0}) describe the same system: we only changed notations and chose a particular vacuum as system's ground state to expand the Lagrangian around. This is why we may speak of a symmetry \emph{spontaneously} broken: no external agent provoked it and the symmetry of the system is still preserved, although ``hidden'' by a particular choice for the ground state.

Taking the limit $\lambda \rightarrow 0$ and plugging $\mathcal{L}_{\phi}$ into the Eulero-Lagrange equations, we obtain the motion equation for fields $\sigma$ and $\theta$ for a one-dimensional diatomic chain along spatial coordinate $x$ as
\begin{eqnarray}
&\left(\frac{\partial^2}{\partial t^2} - c_s^2 \frac{\partial^2}{\partial x^2}\right)  \theta(t,x) =  0 \label{eq:theta} 
    \\
&\left(\frac{\partial^2}{\partial t^2} - c_s^2 \frac{\partial^2}{\partial x^2} + \omega_0^2\right)  \sigma(t,x) =0 
\label{eq:sigma} 
\end{eqnarray}
whose oscillatory solutions $\propto e^{i \left(\omega_{\theta,\sigma}(k) t - k x\right)}$ yield the dispersion relations 
\begin{equation}
\omega_{\theta}^2 = c_s^2  k^2 
\label{eq:theta_dispersion_nonlinear}
\end{equation}
\begin{equation}
\omega_{\sigma}^2 =  \omega_0^2 +  c_s^2  k^2  .
\label{eq:sigma_dispersion_nonlinear}
\end{equation}
They describe respectively a \emph{gapless} and a \emph{gapped} mode of the displacement field $\phi$ (and consequently a gapless and a gapped mode of the ensuing cell's dipole moment, if the solid is ionic), resembling respectively the acoustic and the optical phonon's dispersion law.

From a more formal point of view, we can state that the SSB has generated the real and positively defined mass-like term $\omega_0^2 \sigma^2 $ for the amplitude or Higgs mode $\sigma$ (the optical phonon). From the present approach it is also clear that the optical phonon is not a gauge field, since the oscillation costs at least the energy $\hbar \omega_0$, and it is not possible to change ground state changing $\sigma$ without spending energy. 

In addition, still owing to the spontaneous breaking of a continuous Lagrangian symmetry, a scalar field (here $\theta$) appears and results \emph{massless}, yielding the gapless dispersion relation $\omega_{\theta}^2 =   c_s^2 k^2$. In this limit, it is possible to change the ground state of $\theta$ without spending energy, just operating a gauge transformation along the valley of the Mexican hat (see Figure\,\ref{f:3}(c)). It follows that the field $\theta$ is a pure gauge field that we identify with the \emph{acoustic phonon}, arising as the Goldstone mode associated to the breaking of a continuous symmetry. 

However, it must be stressed that the present approach also describes in an unitary way several nonlinearities, in particular the coupling between optical and acoustic phonons, an argument addressed in section \ref{sec:phonons-coupling}.   

\section{Nonlinearities and phonons-coupling}
\label{sec:phonons-coupling}

An  important outcome of the present SSB approach is the natural emergence of phonon-phonon interactions and self-interactions, ensuing from the terms in $\lambda$ in the expressions of Eq.\,(\ref{eq:quartic_2_0}). The dynamics of $\theta$ alone, described by $\mathcal{L}_{\theta}$ in Eq.\,(\ref{eq:quartic_2_theta}), is the same we obtained gauging the spatial symmetry $T(4)$ in the framework of non-Abelian gauge theories (see Eq.\,(\ref{eq:quartic_R})), revealing a deep link between the two approaches. The term $(\lambda/4) \theta^4$ describes four-phonon processes, like a scattering between two acoustic phonons or a decay of one into three acoustic phonons (all them are impossible in an Abelian theory, like e.g. the electromagnetism). The same argument also applies to optical phonons, whose scattering is ruled by the corresponding term $(\lambda/4) \sigma^4$ in Eq.\,(\ref{eq:quartic_2_sigma}).  

\begin{figure*}[!t]
\centerline{\includegraphics[width=0.5\textwidth]{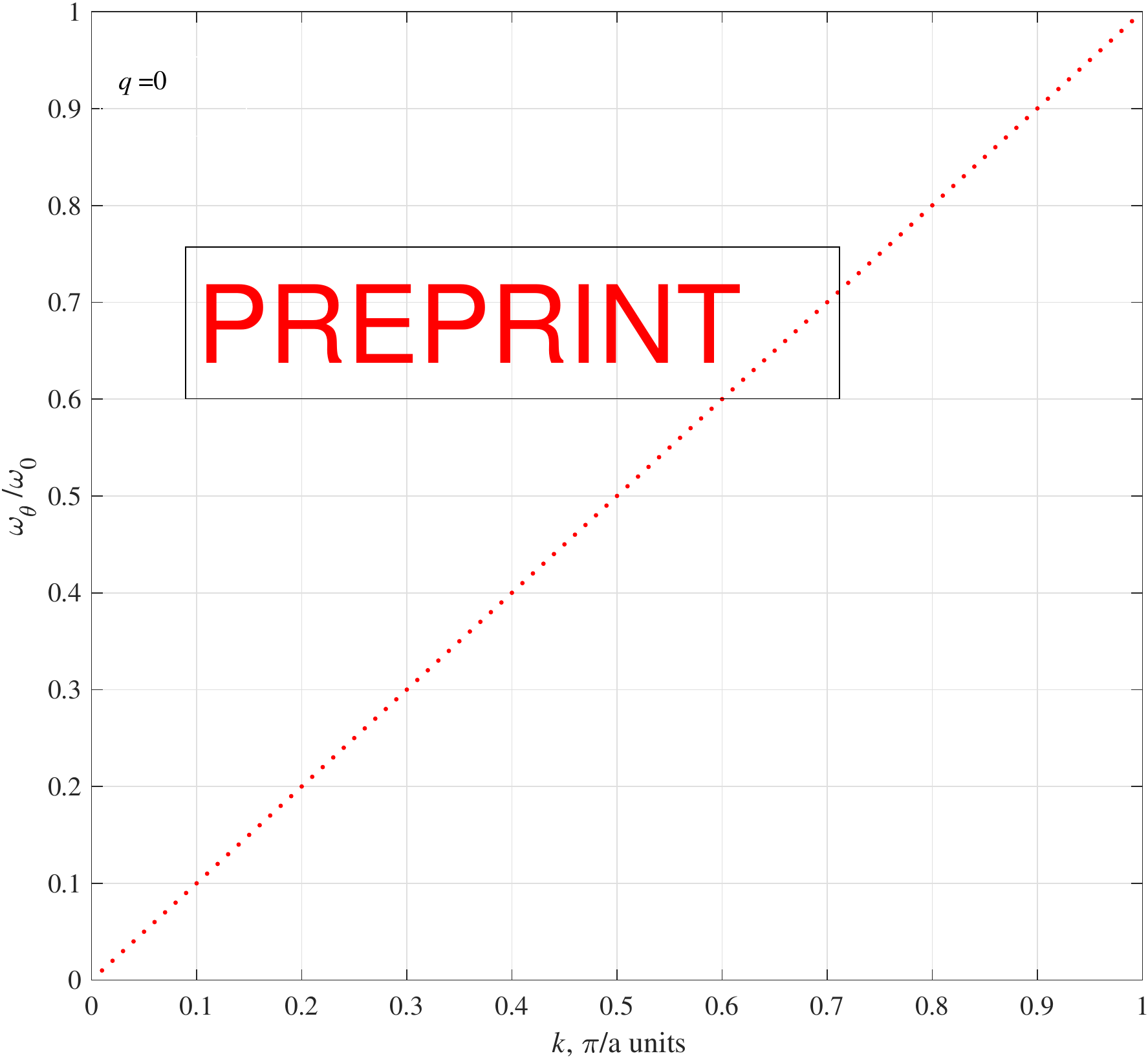}
\includegraphics[width=0.5\textwidth]{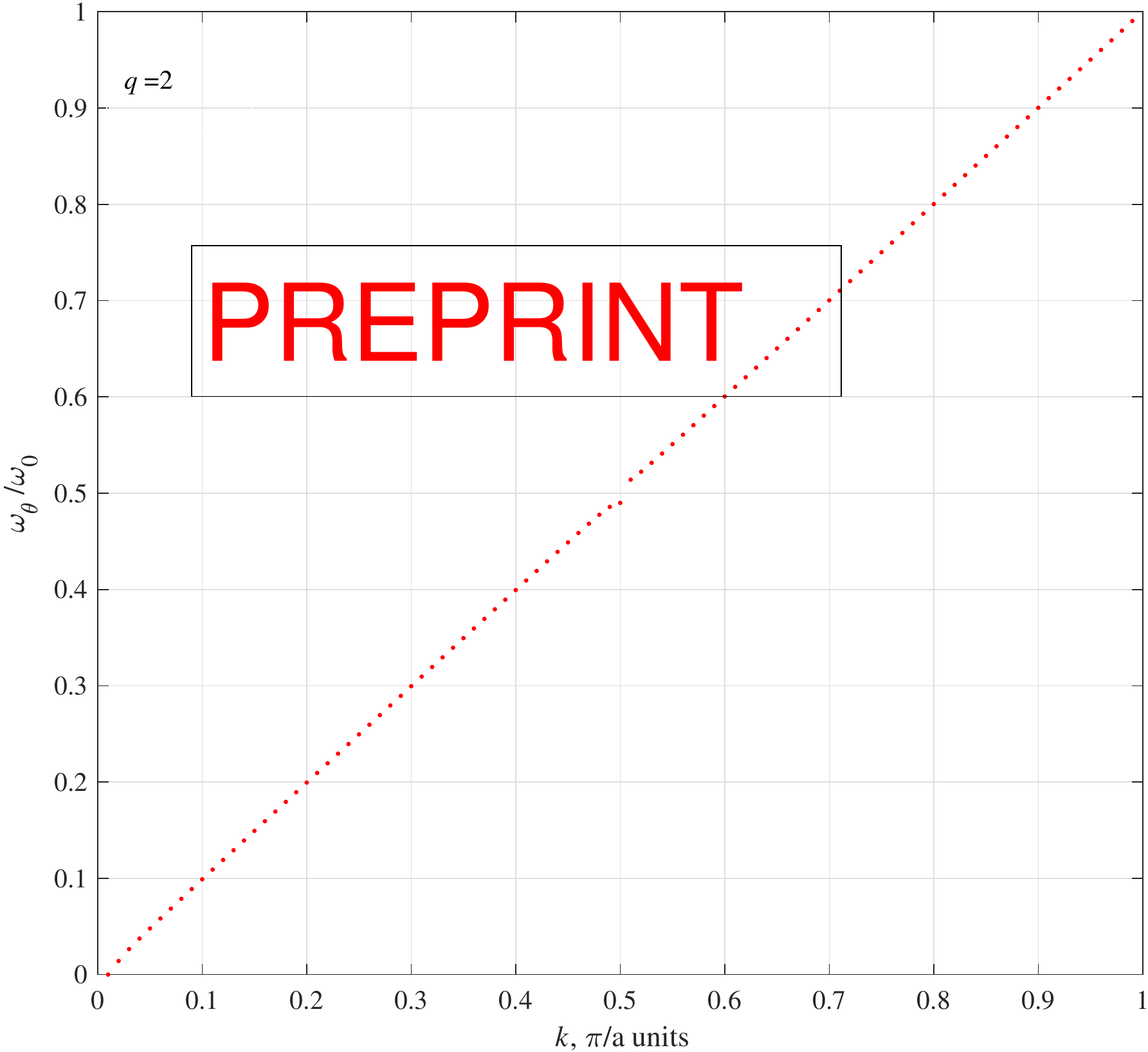}}
\centerline{\includegraphics[width=0.5\textwidth]{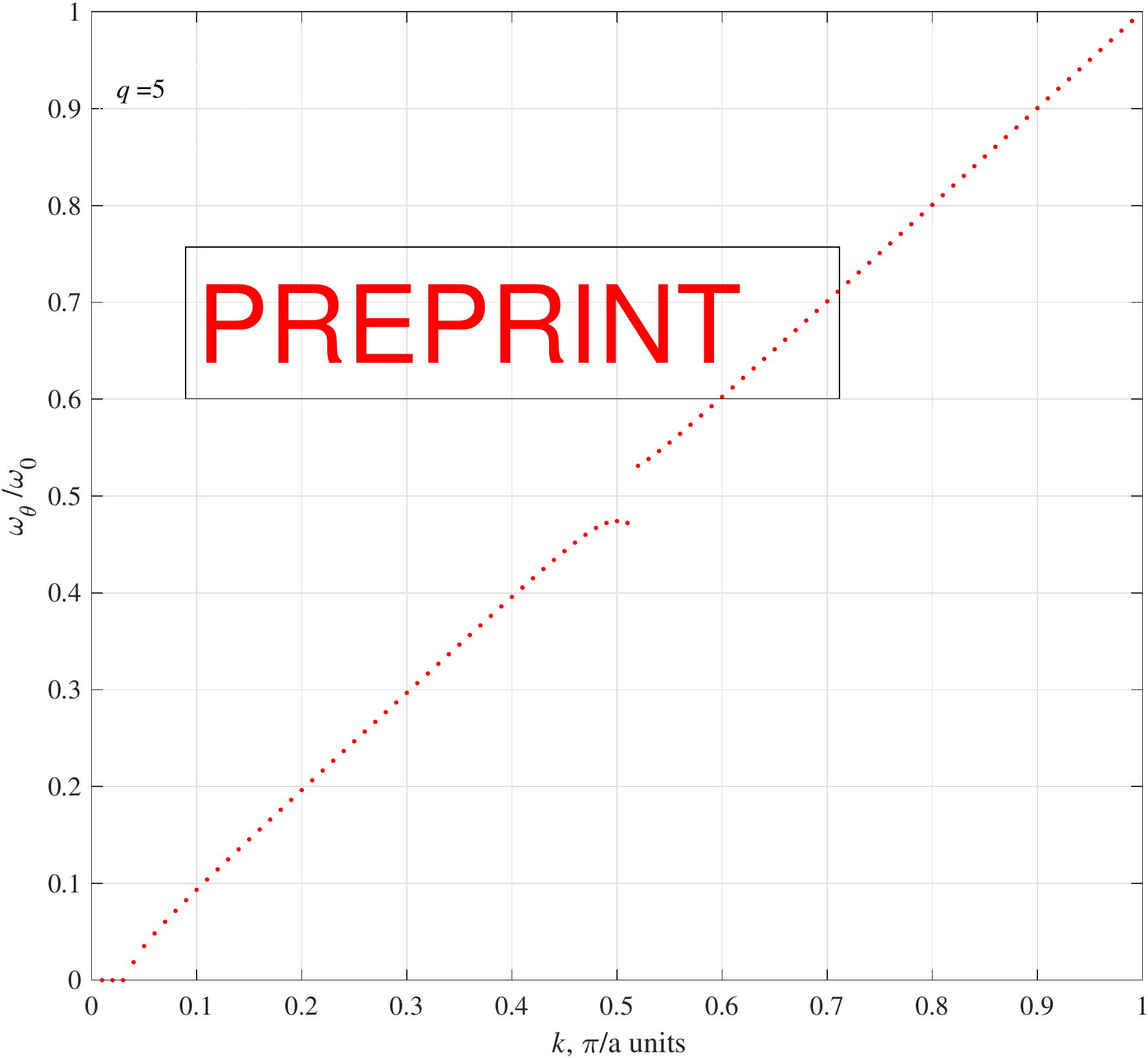}
\includegraphics[width=0.5\textwidth]{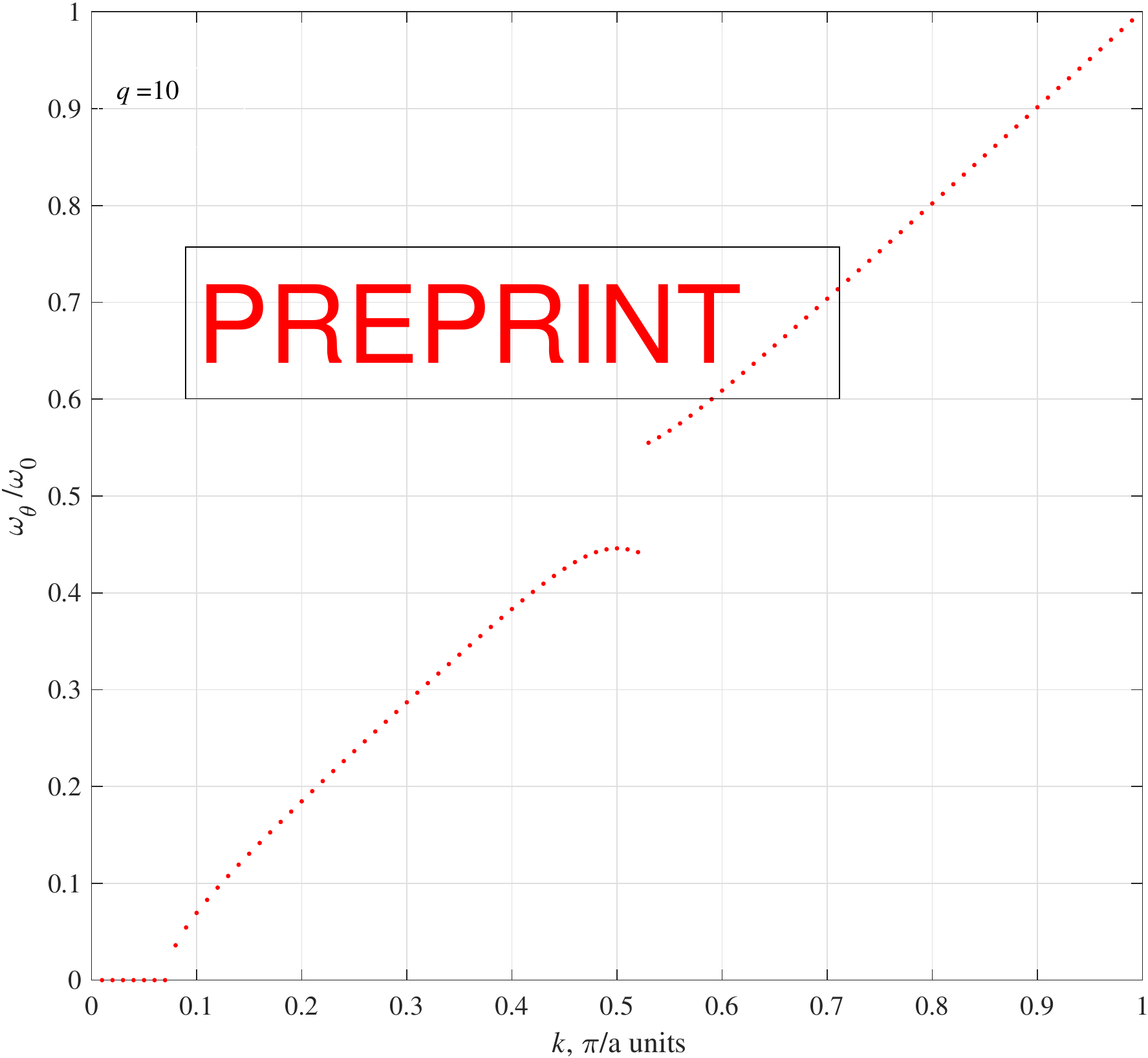}
}
\caption{Dispersion relation for the acoustic phonon: the pure acoustic, unperturbed case $q = 0$, and the effect of the interaction with optical phonons, for $q = 2, 5, 10$, where a mini-gap opens for $\omega_\theta \approx 0.5 \omega_0$. }
\label{f:4}
\end{figure*}

A further important point is the functional form of the Lagrangian $\mathcal{L}_{\sigma-\theta}$, Eq.\,(\ref{eq:quartic_2_sigma_theta}), neglected during the preliminary analysis. It consists of two terms, and it confirms a possible coupling between acoustic and optical phonons, as shown in Refs.\,[\onlinecite{2015Willatzen_PRB,2018Lu_PRB}] for bulk GaAs, GaN, ZnO, MoS$_2$, and for BN monolayers. In Eq.\,(\ref{eq:quartic_2_sigma_theta}), the term proportional to $\sigma \theta^2$ describes the decay of an optical phonon $\sigma$ into two acoustic phonons $\theta$ (or the reverse process), as outlined e.g. in Ref.\,[\onlinecite{1966Klemens_PR}] fitting the theory to silicon. In the same equation, the term proportional to $\theta^2\sigma^2$  describes a scattering between two acoustic and two optical phonons, a four-phonon scattering term whose role has been recently recognized \cite{2017Feng_PRB} in boron arsenide to be responsible for a substantial reduction of its expected thermal conductivity.

At the lowest order in $\lambda$, Eq.\,(\ref{eq:quartic_2_sigma_theta}) provides additional terms in the coupled equations of motion for $\theta$ and $\sigma$, that become
\begin{eqnarray}
&\left(\frac{\partial^2}{\partial t^2} - c_s^2 \frac{\partial^2}{\partial x^2} + \sqrt{2 \lambda } c_s \omega_0 \sigma(t,x)  \right)  \theta(t,x) =  0  
\label{eq:theta_equation_nonlinear} 
 \\
&\left(\frac{\partial^2}{\partial t^2} - c_s^2 \frac{\partial^2}{\partial x^2} + \omega_0^2\right)  \sigma(t,x) =0 .
\label{eq:sigma_equation_nonlinear}
\end{eqnarray}

The equation for $\sigma(t,x)$ is the same Eq.\,(\ref{eq:sigma}): without loosing generality, we can write its oscillatory solution as $\sigma(t,x) \propto \cos\left(\omega_{\sigma} t - k x\right)$ and plug it in Eq.\,(\ref{eq:theta_equation_nonlinear}). Approximating $\omega_{\sigma} \approx \omega_0$, separating the variables as $\theta(t,x) = \theta_t(t) \theta_x(x)$, and supposing $\theta_x(x) \propto \exp\left(i k x\right)$, the equation of motion for $\theta_t(t)$ is 
\begin{equation}
\frac{d^2 \theta_t(t)}{dt^2} + \left(c_s^2 k_n^2 + q \,\omega_0  \cos\left(\omega_0 t \right)  \right)  \theta_t(t) =  0
\label{eq:theta_equation_Mathieu}
\end{equation}
where $q = \sqrt{2 \lambda}$. For simplicity, in the calculations it is also possible to set $c_s=1$. The phonon wavevector takes the discrete values $k_n = n \pi /(N a)$, where $n = 1, ...N$, and $N$ is the number of atoms in the chain. Eq.\,(\ref{eq:theta_equation_Mathieu}) can be treated as a Mathieu's equation \cite{1972Abramowitz,1953Morse,2009Coisson,2016Parra_JPCS}, which describes time-dependent harmonic oscillators perturbed by a periodic load (the term in the cosine) representing the interaction with the optical phonon. For $\lambda=0$ the eigenfrequencies are $\omega_\theta = c_s k_n$, recovering the standard acoustic phonon dispersion law in the long wavelength limit. 

In the general case $q \neq 0$, we treated the perturbation induced on the cell by the \emph{optical} oscillation as a boundary value problem according to Ref.\,[\onlinecite{1987Ascher}]. In short, Eq.\,(\ref{eq:theta_equation_Mathieu}) was written as $\theta_t'' + \left(b_n + q \,\omega_0 \cos\left(\omega_0 t \right)  \right)  \theta_t$, where $b_n$ is an unknown parameter, to be regarded as a discrete eigenvalue. Defining the domain as $t \in[0, T]$ and imposing as boundary conditions $\theta_t(0) = 1$, and $\theta_t'(0) = \theta_t'(T) = 0$ (cosine-like solutions), the eigenvalues $b_n$ were found iteratively, starting from a guess solution $\theta_t=\cos(n \pi t/T)$ and a guess value $b_n$. The convergence is fast, and the obtained parameters $b_n$ can be regarded as the discrete eigenvalues $\omega_\theta^2$ associated to eigenfunctions $\theta_t$. 

In order to see how the optical-acoustic interaction \emph{qualitatively} works, since the energy of optical phonons is usually higher than acoustics', it makes sense to set $\omega_0 = c_s k_N$. It is interesting that for $q \ne 0$ the present formulation predicts the existence of mini-gaps opening inside the Brillouin zone, for $\omega_\theta \approx 0.5 \omega_0$, as visible in Figure\,\ref{f:4}. This is a feature typical of Mathieu's equation: a periodic modulation induces a Bragg-like condition that prevents the phonon propagation. Similar behavior was predicted and experimentally observed by  Raman spectroscopy in superlattices \cite{1987Tamura_PRB,1989Calle_SSC,2016Shinokita_PRL,2007Bruchhausen_JPCS}: in both cases, there is a periodic modulation (caused by the optical mode according to the present work, or caused by the superlattice in the cited cases) that induces a Bragg condition which prevents the phonon propagation when its frequency is around half of the modulation frequency. Furthermore, piezoelectrical coupling of optical and acoustic phonon modes was found also in zinc-blende GaAs quantum-well slabs \cite{2015Willatzen_PRB,2017Duggen_PRB} and theoretically calculated for wurzites in the Lagrangian formalism in Ref.\,[\onlinecite{2017Datta_ICN}], confirming by different approaches the existence of the phenomenology described in the present work.

\section{Conclusions}
\label{sec:conclusions}

We investigated in depth the emergence of acoustic phonons in crystalline solids, both as Goldstone and as gauge bosons. The acoustic phonon was previously acknowledged as the gauge boson appearing when the electron's Lagrangian in a crystalline solid is requested to be locally invariant with respect to the group of spatial translations, whose generators appear broken due to the lattice itself. 

However, the gauging of the translations group is not trivial: in order to better explain the context, first we reviewed the differences between the gauging of spatial (external) and internal symmetries, as in Yang-Mills' theories. Then, exploiting the mathematical similarities between the sonic (or acoustic) metric and the Lorentzian metric of the ordinary space-time, we gauged the translations group in solid crystals following the same tetrad (or vierbein) formalism employed to express the General Relativity as a gauge theory with space-time translations as structure group. This allowed to compare the dynamical role of the crystal elasticity tensor in solid state physics and the Ricci's tensor in General Relativity at a more formal and rigorous level than before. The subsequent analysis showed that, beyond the linear limit, the translations group's non-Abelianity may generate a gap in the frequency dispersion relation of the acoustic phonon, as if a mass-like term were present in its free field Lagrangian, although this is not the case. 

In order to better investigate the latter point, we considered a different scenario, driving attention to the problem of understanding the nature of the \emph{optical phonons} in standard solids. We showed that both the acoustic
and optical phonon emerge respectively as the gapless Goldstone (phase) and the gapped Higgs (amplitude) fluctuation mode of an order parameter arising from the spontaneous breaking of a global symmetry, without invoking the gauge principle. In greater detail, following the linear-sigma model approach the Higgs' mechanism is shown to generate a massive amplitude (Higgs) mode that we identify with the optical phonon, with gapped frequency dispersion relation due to the mass-like term. At the same time, the Higgs' mechanism generates a massless phase (Goldstone) mode, the acoustic phonon. A frequency-gap only appears in the strong nonlinear regime, and it is due to an anharmonic term, the same that arises from the gauging of the spatial translations group, an approach which did not provide any description of the optical phonon, though.

Although the SSB is a well known phenomenon, it turns out that it is able to describe not only novel aspects of material science (like e.g. Cooper pair plasmon modes in superconductors layered cuprates and thin films \cite{1993Cote_PRB,2011Podolski_PRB}), but also an old acquaintance of the solid state physics, as the acoustic and optical phonons in crystals. In particular, the Higgs mechanism describes all the phonon-phonon interactions, including a possible perturbation on the acoustic phonon's frequency dispersion relation induced by the eventual optical phonon, a peculiar behavior not described so far, at the best of author's knowledge, able to produce unexpected mini-gaps inside the Brillouin zone.

%

\providecommand{\WileyBibTextsc}{}
\let\textsc\WileyBibTextsc
\providecommand{\othercit}{}
\providecommand{\jr}[1]{#1}
\providecommand{\etal}{~et~al.}

\end{document}